# IONIZATION RATE AND PLASMA DYNAMICS AT 3.9 MICRON FEMTOSECOND PHOTOIONIZATION OF AIR

Adam Patel[1], Claudia Gollner[2], Rokas Jutas[2], Valentina Shumakova[2], Mikhail N. Shneider,[3] Audrius Pugzlys[2], Andrius Baltuska[2], and Alexey Shashurin[1]

[1] *School of Aeronautics and Astronautics, Purdue University, West Lafayette, IN, USA*
[2] *Photonics Institute, Vienna University of Technology, Vienna, Austria*
[3] *Mechanical and Aerospace Engineering, Princeton University, Princeton, NJ, USA*

## Abstract

The introduction of mid-IR optical parametric chirped pulse amplifiers (OPCPAs) has catalyzed interest in multi-millijoule, infrared femtosecond pulse-based filamentation. As tunneling ionization is a fundamental first stage in these high-intensity laser-matter interactions, characterizing the process is critical to understand derivative topical studies on femtosecond filamentation and self-focusing. Here, we report first direct nonintrusive measurements of total electron count and electron number densities generated at 3.9 μm femtosecond mid-infrared tunneling ionization of atmospheric air using constructive-elastic microwave scattering. Subsequently, we determine photoionization rates to be in the range of $5.0 \times 10^8 - 6.1 \times 10^9$ s$^{-1}$ for radiation intensities of $1.3 \times 10^{13} - 1.9 \times 10^{14}$ W/cm$^2$, respectively. The proposed approach paves the wave to precisely tabulate photoionization rates in mid-IR for broad range of intensities and gas types and to study plasma dynamics at mid-IR filamentation.

**Corresponding author:** Alexey Shashurin ashashur@purdue.edu

*Introduction.* - Strong field ionization of gases by intense femtosecond laser pulses enables a plethora of frontier fundamental research fields in nonlinear optics, such as high-order harmonic generation (HHG) and attosecond science,[1] femtosecond filamentation,[2,3] standoff nonlinear spectroscopy of atmosphere,[4] etc. In contrast to optical breakdown triggered by picosecond (and longer) laser pulses or microwave discharges (where Joule heating of photoelectrons and avalanche ionization play a major role in evolution of the plasma, density, temperature, and energy/momentum distribution of photoelectrons), plasmas generated by femtosecond pulses in atmospheric pressure gases are fully determined by the fundamental quantum process of strong field ionization. Very recent progress in the development of high-energy femtosecond laser systems in the mid-infrared (mid-IR) spectral range has opened new regimes and perspectives in strong field physics and femtosecond filamentation.[5,6,7,8,9] Specifically, these refer to the increased propagation range of the mid-IR laser radiation through the atmosphere in comparison to their traditional near-IR counterparts and, therefore, enable a variety of applications including remote sensing of the atmosphere, free-space communications, directed energy applications, and laser propulsion for deep-space travel.[4,5,6,7,8,10,11]

Accurate knowledge of the plasma properties, generated by intense ultrashort mid-IR laser pulses, and their dependence on laser pulse parameters has critical importance for understanding physics and the precision of numerical simulations of HHG and generation of attosecond pulses, laser wake-field acceleration,[12] femtosecond filamentation, and related phenomena like standoff terahertz emission generation,[13] laser-controlled weather phenomena,[14] standoff lasing,[4,15] etc. The dependence of ionization rate on laser intensity and wavelength for atomic gases was theoretically described in seminal works.[16,17,18] In the latest development of these theoretical models, it was shown that the ionization probability as well as the momentum and energy distributions of photoelectrons are strongly affected by the laser wavelength and polarization.[19] However, even for atomic gases, there are no absolute measurements of the plasma parameters and ionization rates generated by the laser pulse in mid-IR, and experiments conducted thus far are mostly in a narrow spectral range of 0.8-1 μm with near-IR lasers, providing semi-empirical agreement with theoretical predictions.[20,21] The status quo in molecular gases, demonstrating additional strong dependence of ionization on molecular orientation and internal ultrafast dynamics,[22,23] is even worse and, despite extensive theoretical efforts to extend atomic strong field ionization models to molecules,[24] simulations and experiments currently rely on empirical dependences retrieved for a limited range of intensities and for particular laser wavelengths.[25,26,27]

Measurements of absolute electron numbers generated in gases by intense femtosecond laser pulses and corresponding photoionization rates are very challenging. For relatively high plasma densities, $n_e \geq 10^{16}$ cm$^{-3}$ methods of optical interferometry or shadowgraphy can be employed.[28,29] However, the ionization rate is predicted to drop with increasing laser wavelength,[5] and methods for electron density measurements in mid-IR filaments below the limit of $n_e \leq 10^{16}$ cm$^{-3}$ are required. A number of semi-empirical and/or indirect methods was proposed, including ion current measurement by time-of-flight mass spectrometer, scattering of THz radiation, measurements of capacitive response times, picosecond avalanche electron multiplication, and attenuation of $TE_{10}$ mode in rectangular waveguide; however, these methods yield relative and/or indirect measurements associated with ambiguous interpretation.[9,25,28,29,30,31,32,33] Direct counting of photoionized electrons based on picosecond avalanche amplification of individual electrons was reported recently and the method is well-suited for very low ionization degrees $\leq 10^{-10}$.[34]

Recently, coherent microwave scattering (CMS) was demonstrated as an effective diagnostic technique for the measurement of photoionization rates and temporally-resolved electron number densities at laser-induced photoionization in UV and near-IR spectral ranges.[35,36,37] State-of-the-art CMS systems are associated with high sensitivity and minimal detectable number of electrons in microplasma volume $\sim 10^8$. In this work, we expand CMS application to measure the total electron count and electron number density produced by focusing a 3.9 μm femtosecond laser pulse in air and subsequently determine tunnelling ionization rates of ambient air in the mid-IR spectral band.

*Experimental Details and Methodology.* - The experiments were conducted at the Ultrafast Laser Group of the Photonics Institute at TU Wien (Austria) with the unique 3.9 μm 30 mJ mid-IR femtosecond OPCPA.[7,38,39] The coherent microwave scattering (CMS) measurement system was provided by Purdue University (USA) to conduct these experiments. The experimental setup including the femtosecond mid-infrared laser and CMS system used in this work is schematically shown in Figure 1.

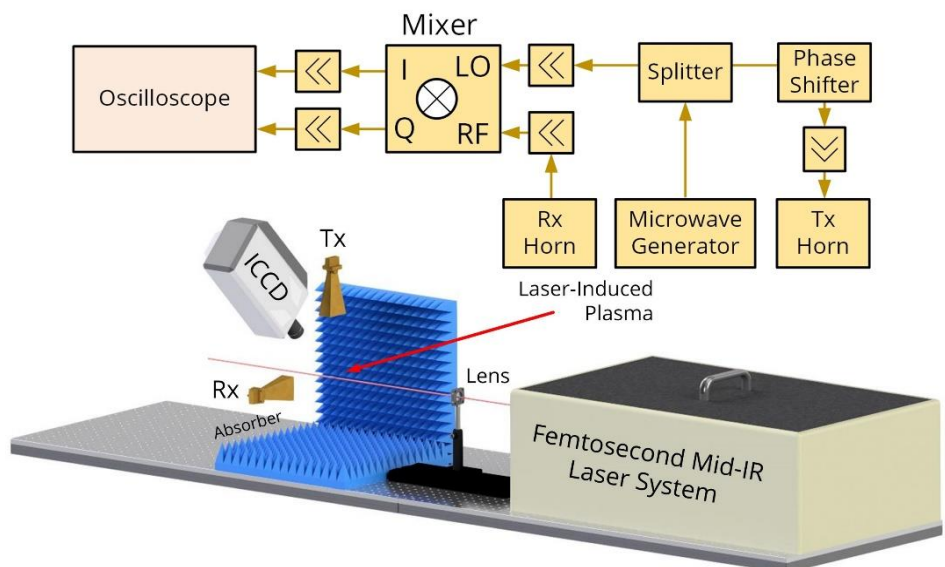

**Figure 1. Schematics of the experimental setup containing the 3.9 μm femtosecond laser and coherent microwave scattering system.**

Measurements of the total electron count in the laser-induced plasma volume ($N_e$) and electron number density ($n_e$) were conducted using a homodyne-type CMS system operating at 11 GHz. Incident microwave radiation (~8 W, Tx horn) was linearly polarized along the plasma channel orientation, and the signal scattered off the plasma volume was received by a microwave horn (Rx horn), amplified, and detected using an I/Q mixer. All components of the microwave system operated in the linear regime to ensure that the measured response is proportional to the amplitude of the scattered signal. The temporal resolution of the CMS system was about 1 ns, limited by 1 GHz bandwidth LeCroy WaveRunner 104Xi oscilloscope used in the circuit. The CMS system was calibrated using Teflon dielectric scatterers and attribution of the measured scattering signal $U_S$ to $N_e$ was done via the following equation:

$$U_S = \begin{cases} A \dfrac{e^2}{m\nu_m} N_e & - \text{ laser} - \text{induced plasma} \\ A V_D \varepsilon_0 (\varepsilon_D - 1)\omega & - \text{ dielectric scatterer} \end{cases} \quad \textbf{Eq. 1}$$

where $A$ – the calibration coefficient of the CMS system (evaluated from the lower equation), $V_D$ and $\varepsilon_D$ – the volume and the dielectric constant of the Teflon dielectric scatterer, $\varepsilon_0$ - the dielectric permittivity of vacuum, $\nu_m$ – the electron-gas collisional frequency, and $\omega$ – the circular frequency of the microwave radiation. The laser-induced plasma was imaged using an ICCD camera to observe its evolution with laser pulse energy and determine plasma volume $V_p$. Spatially-averaged electron number density was determined as $n_e = \dfrac{N_e}{V_p}$. More details on the CMS technique can be found elsewhere.[35,36,40]

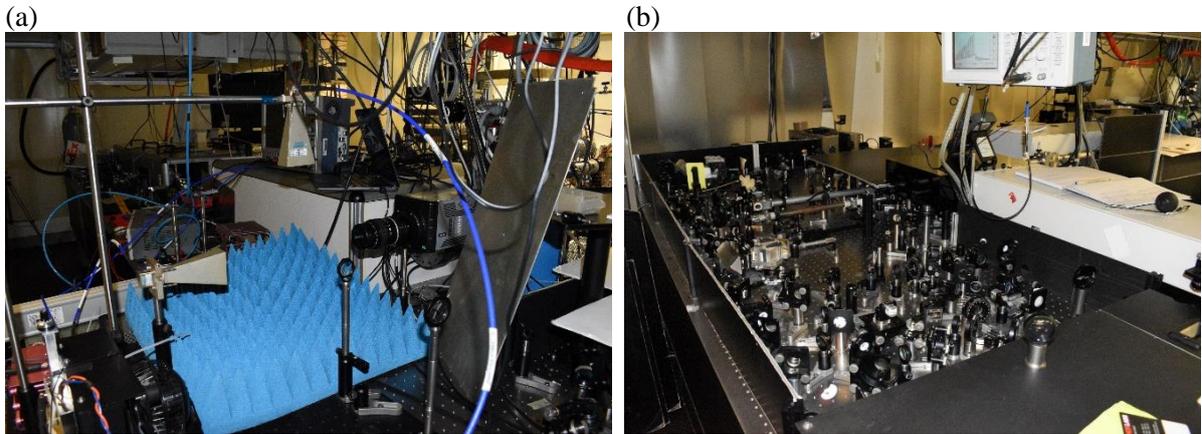

**Figure 2. (a)** Photograph of CMS system aligned with the beam waist location. **(b)** Photograph of the 3.9 μm 30 mJ femtosecond laser system.

*Results and Discussion.* - In the experiments, linearly-polarized 3.9 μm pulses were focused in air by a 150 mm focal length CaF2 lens. Femtosecond laser pulses were characterized by second-harmonic generation frequency resolved optical-gating (SHG FROG) method, beam-waist was measured by a knife-edge technique (conducted with a strongly attenuated laser beam), and spatial distribution along the propagation was monitored with a Spiricon Pyrocam III beam profiler. The results of these measurements are shown in Figure 3. SHG FROG measurement reveals spectral FWHM – 302.9 nm and temporal FWHM – $\tau_{FWHM}$=117.7 fs. Before focusing, the 3.9 μm beam was slightly elliptical with the beam diameters in x and y directions at FWHM level being 3.98 mm and 4.64 mm, respectively. The beam waist was evaluated to be $w_0$=95.85 μm. These quantities can be regarded as invariant for the experimental conditions encountered in this work.

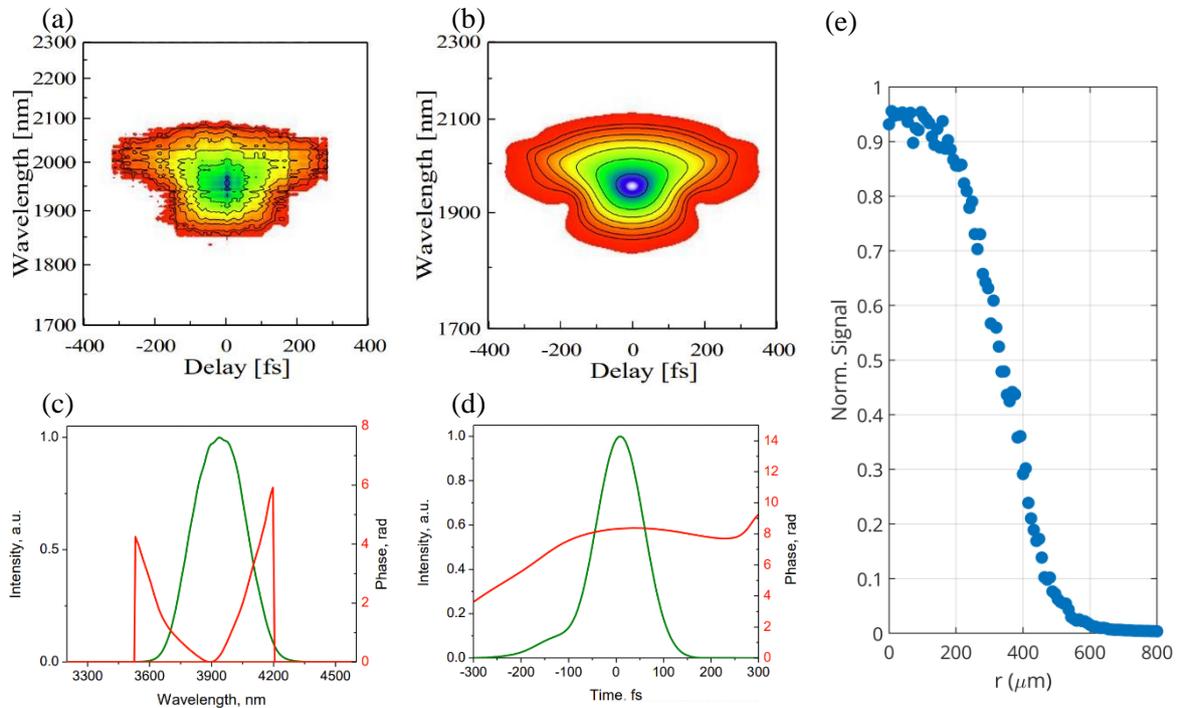

**Figure 3**. **Characteristics of the femtosecond pulse for $E_L$ = 0.2 mJ.** **(a)** Measured SHG FROG. **(b)** Retrieved SHG FROG. **(c)** Spectral composition and phase. **(d)** Temporal profile and phase. **(e)** Knife-edge beam waist measurements for f = 150 mm.

Representative line-of-sight ICCD images (gate width 100 ns) of laser-induced plasmas in ambient air and spatial beam-profiling acquired 15 cm behind the focal plane are depicted in Figure 4 for various laser pulse energies. As observed, and consistent with characteristics of the laser / focusing arrangement, encountered plasma channels are rather short, on the order of a millimeter. One can conclude that beam propagation for the tested irradiating energies $E_L$ up to ~3.5 mJ can be regarded as nearly linear. This is justified by (1) relative invariance of the laser beam profile and (2) symmetry (longitudinal and transversal) of plasma geometric features up to $E_L$ ~ 2 mJ as well as by a fairly weak shift of the focal point in the focusing lens direction. Slight onset of the nonlinear optical effects is encountered at the upper edge of tested energy range $E_L$ ~ 2-3.5 mJ. Note that Kerr focusing and plasma defocusing are expected to have comparable magnitude of contributions to the refractive index (n) for the range of tested laser intensities ($I_L$) and electron number ($n_e$) densities measured below; namely, $|\Delta n| \approx 5 \times 10^{-5}$ for $I_L=10^{18}$ W/m$^2$ and $n_e=10^{16}$ cm$^{-3}$. Aside, the plasma channel can be regarded as an oblate spheroid with semi-axis estimates of $\mathfrak{a} = \mathfrak{b} = 100$ μm (consistent with knife-edge measurements) and c = 0.75 mm. These dimensions suggest plasma volume $V_p=3.14\times10^{-11}$ m$^{-3}$ and somewhat large depolarization factors (governing magnitude of the microwave field screening inside the plasma spheroid due to accumulation of charges on the spheroid ends) on the order of $\xi \approx 0.03$.[35,36] This, coupled with skin-depth considerations, largely restricts the measurable $n_e$ up to about $10^{22}$ m$^{-3}$ before a deviation from ideal collisional coherent microwave scattering behavior (and still below the threshold for significant electron-ion Coulomb contributions to $\nu_m$). Note that longitudinal and transversal plasma boundaries determined using ICCD imaging were associated with intensities of about 10% and about 90% of the peak intensity, respectively (for the Gaussian beam with $w_0$=95.85 μm and Rayleigh length $z_R$=4.1 mm). (Even though this result is consistent with our previous work at near-IR, it should be specifically considered/interpreted in the future)[36].

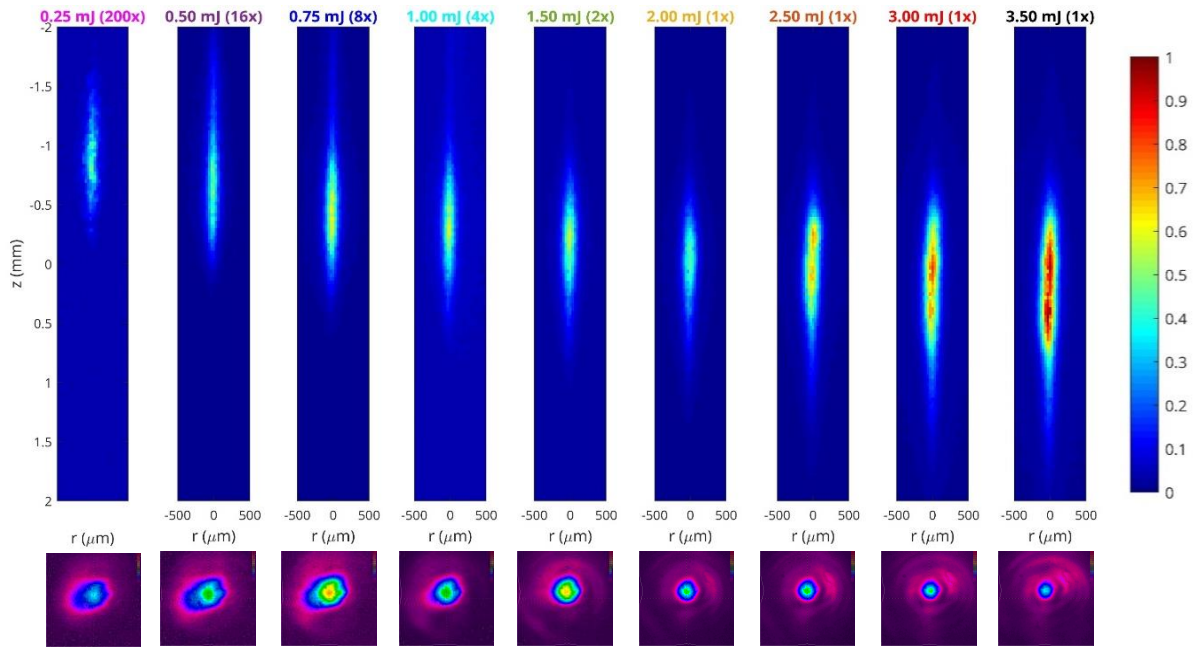

**Figure 4. ICCD images (exposure time 100 ns) and spatial beam profiling (acquired 15 cm behind the focal plane) associated with femtosecond mid-infrared photoionization in atmospheric air for irradiating energies E$_L$ up to 3.5 mJ and f = 150 mm lens.**

Temporal evolution of total electron numbers measured by the CMS is plotted in Figure 5(a) for various E$_L$ (line-colors refer to the headers in Figure 4). $N_e$ was extracted from the measured scattered signal $U_s$ under the assumption of a constant collisional frequency $\nu_m = 2 \cdot 10^{12}$ s$^{-1}$ corresponding to an electron temperature of $T_e \sim 1$ eV. This is consistent with quick $T_e$ collapse from the range corresponding to quiver energy ($E_\sim = e^2 I_L/2c\varepsilon_0 m_e \omega_L^2 \approx 10 - 100$ eV) to an approximately single eV/sub-eV range on the timescale of <<1 ns followed by fairly slow evolution during several next ns.[28] Note that during this quick $T_e$-collapse phase the electrons can produce post-pulse ionization of air and, therefore, reported $N_e$-values might be slightly overestimated (about factor of 2-3). Additionally, observed CMS measurements unrealistically reflect (underestimate) the initial growth stage of $N_e$ temporal dynamics due to limited temporal resolution of the detecting system of about 1 ns (oscilloscope with 1 GHz bandwidth was used). The decay is governed by dissociative recombination ($X_2^+ + e \rightarrow X + X$, where $X$ is $N$ or $O$) and three-body attachment to oxygen ($e + O_2 + X_2 \rightarrow O_2^- + X_2$, where $X$ is $N$ or $O$) and adequately resolved in time by the utilized CMS system.

Temporal evolution of the relative phase of detected microwave radiation measured by the CMS system is plotted in Figure 5(b). One can see that, for all tested pulse energies, the phase quickly collapsed to a constant value and remained unchanged during the subsequent electron decay which is indicative of a collisional scattering regime. This validates the mathematical formulation used to evaluate $N_e$ from the measured scattered signal $U_s$ (upper Eq. 1). Note that pulse energies $E_L$>2mJ might indicate a mixed collisional-Rayleigh scattering regime during first nanosecond after the laser pulse.

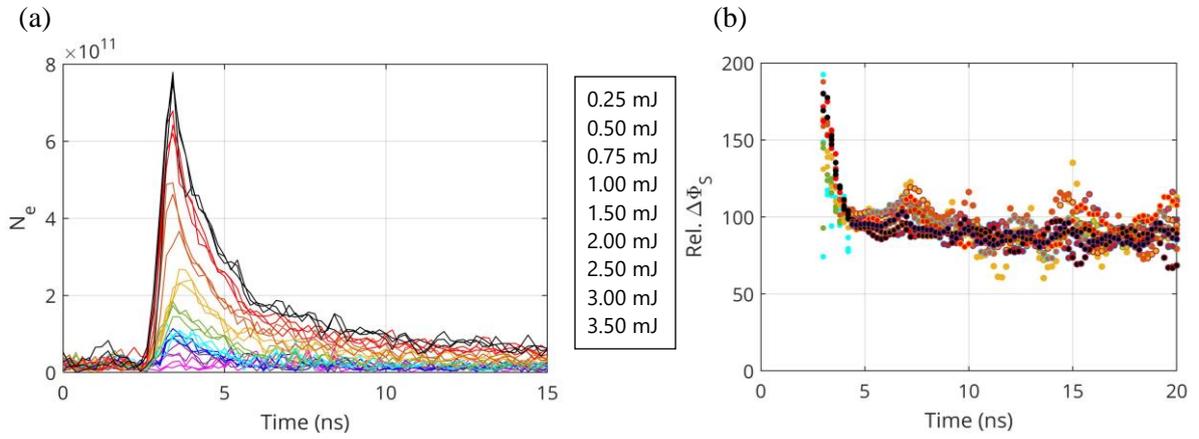

**Figure 5.** **(a)** Temporal decay of $N_e$ measured via CMS for various $E_L$. Line coloring corresponds to the headers in Figure 4. **(b)** Associated relative phase waveforms.

Spatially averaged electron number density ($n_e$) and photoionization rate ($\nu_{PI}$) are plotted as a function of laser intensity in Figure 6. $n_e$ was determined as $n_e = \frac{N_e}{V_p}$ and $\nu_{PI}$ was subsequently calculated from the simple relation $n_e = n_g \nu_{PI} \tau_{FWHM}$, where $n_g$ is air number density ($n_g$=2.5×10$^{25}$ m$^{-3}$). Laser intensities indicated on the horizontal axis were calculated assuming Gaussian beam intensity distribution in spatial and temporal domains in vicinity of the waist using $I_L = \frac{E_L}{\frac{1}{2}\pi w_0^2} \frac{2\sqrt{\ln 2}}{\tau_{FWHM}\sqrt{\pi}}$. Note that assumption of Gaussian intensity distribution is based on experimentally validated linear beam propagation with weak contribution of nonlinear optical effects as detailed above. It was observed that total electron count $N_e$ scaled with laser energy $E_L$ as $N_e \propto E_L^{(1.5-1.7)}$; however, determination of effective plasma volume based on this scaling law (developed and utilized in our previous works)[36,37] was not pursued in this work as it was beyond the accuracy of the current dataset associated with limited size and substantial measurement errors.

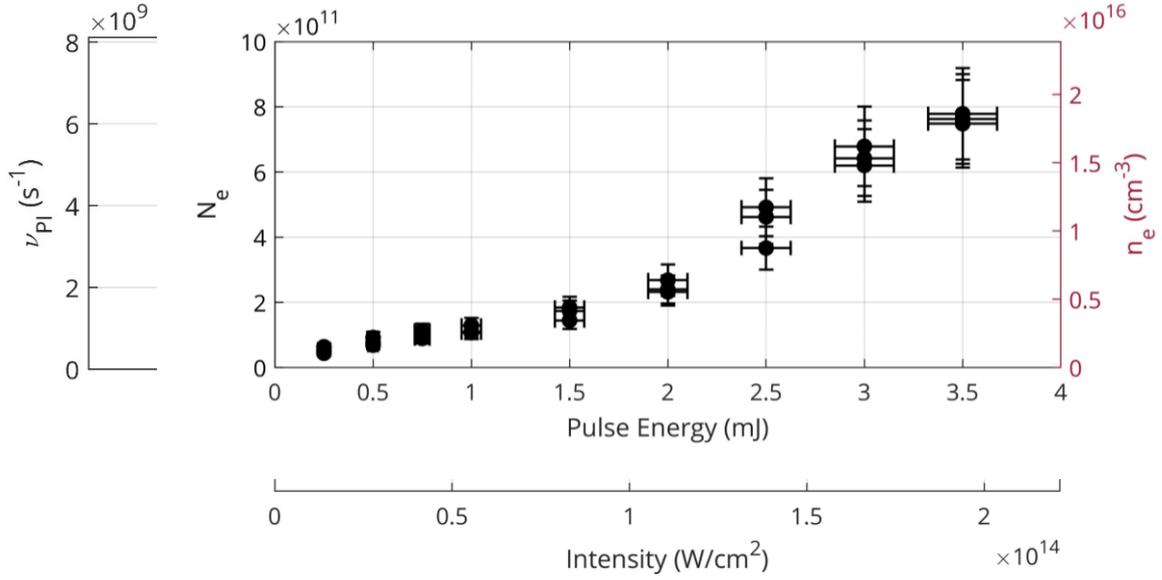

**Figure 6. Experimentally-measured total electron numbers, electron number densities, and photoionization rates as a function of pulse energy / peak intensity.**

In summary, photoionization rates at 3.9 μm wavelength were experimentally determined for the range of radiation intensities $1.3 \times 10^{13}$ – $1.9 \times 10^{14}$ W/cm² using 3.9 μm femtosecond laser and absolutely calibrated coherent microwave scattering technique. Several important modifications can be considered in the future to improve accuracy of $\nu_{PI}$ determination and extend the intensity range. First, more rigor procedure for identifying effective plasma volume via deriving exact power-law scaling of $N_e$ with $E_L$, extrapolating the same scaling locally on $n_e$ vs. $I_L$ dependence, and conducting corresponding integrations similar to the methodology developed in our recent works.[36,37] This would further introduce a correction to the effective plasma volume $V_p$ and, correspondingly, to photoionization rate $\nu_{PI}$. Second, experiments in pure gases should be conducted to identify individual contributions of various gaseous species containing in the atmospheric air ($O_2$, $N_2$, etc.) to the overall photoionization rate. Note that in current work we calculated $\nu_{PI}$ based on the total air density $n_g = 2.5 \times 10^{25}$ m⁻³ so that higher gas-specific photoionization rates (e.g., $\nu_{O_2}$, $\nu_{N_2}$) should be expected ($\nu_{PI} = \nu_{O_2}\chi_{O_2} + \nu_{N_2}\chi_{N_2}$, where $\chi_{O_2}$ and $\chi_{N_2}$ are mole fractions of oxygen and nitrogen in air). Photoionizations rates measured in this work coincide with the lower bound of the corresponding range predicted by various theoretical models developed in the past ($10^9$-$10^{13}$ s⁻¹ for the radiation intensity ~$10^{14}$ W/cm² with intensity scaling varying in the range $I_L^{(1-5)}$)[25]. Note that we compare the photoionizations rates measured in this work at 3.9 μm with that reported previously at 800 nm as similar rates in the tunneling regime are expected (largely governed by the amplitude of electric field rather than by the photon energy/wavelength determining the multiphoton regime). Third, conducting experiments in collisionless $v_m$-independent Thomson scattering regime at low pressures ($p \leq 1$ Torr) would allow to substantially extended the range of intensities at which $v_{PI}$ can be evaluated by delaying the onset of nonlinear optical phenomena to higher laser intensities.[35] This refers to the fact that nonlinear optical phenomena including Kerr and plasma nonlinear terms in the refractive index reduce proportionally with a decrease in pressure ($n_2 \propto n_g \propto p$, $\omega_p^2 \propto n_e \propto n_g \propto p$). Forth, expanding bandwidth of the CMS measurement system is recommended to accurately capture temporal dynamics of $N_e$ and fully resolve $N_e$ peak value.

*Conclusions.* - In this work, total electron count, electron number densities, and photoionization rates were reported for mid-infrared femtosecond tunneling photoionization at 3.9 μm in atmospheric air using absolutely calibrated coherent microwave scattering technique. Utilization of atmospheric pressure limits the highest intensity at which a photoionization rate can be evaluated due to onset of Kerr focusing. To extend this limit to higher intensities, the experiments at low pressures ($p \leq 1$ Torr) in collisionless Thomson scattering regime should be conducted.

*Acknowledgements.* - Authors would like to thank D. Kartashov and M. N. Slipchenko for valuable discussions. This work was supported by the National Science Foundation (Grant No. 1903415 and No. 1903360).